\documentclass[12pt]{article}

\usepackage{amsfonts}
\usepackage{graphicx}

\newcommand{\un}{{\mathbb I}}
\newcommand{\ra}{\rightarrow}

\newcommand{\bra}{\langle} 
\newcommand{\ket}{\rangle}
\newcommand{\upket}{|\!\uparrow\rangle}
\newcommand{\downket}{|\!\downarrow\rangle}
\newcommand{\upbra}{\langle\uparrow\!|}
\newcommand{\downbra}{\langle\downarrow\!|}
\renewcommand{\i}{{\rm i}}
\newcommand{\E}{{\mathbb E}}

\newcommand{\be}{\begin{equation}}
\newcommand{\ee}{\end{equation}}
\newcommand{\bea}{\begin{eqnarray}}
\newcommand{\eea}{\end{eqnarray}}
\newcommand{\cx}{{\mathbb C}}
\newcommand{\cz}{{\mathbb Z}}
\newcommand{\ch}{{\cal H}}

\newcommand{\ffi}{\varphi}

\newcommand{\grintl}{[\kern-.18em [}
\newcommand{\grintr}{]\kern-.18em ]}

\newcounter{resultcounter}[section]

\newtheorem{thm}[resultcounter]{Theorem}
\newtheorem{lem}[resultcounter]{Lemma}
\newtheorem{prop}[resultcounter]{Proposition}
\newtheorem{cor}[resultcounter]{Corollary}
\newtheorem{definition}[resultcounter]{Definition}
\newtheorem{rem}[resultcounter]{Remark}

\def\bed{\begin{definition}}
\def\eed{\end{definition}}

\newcommand{\R}{{\mathbb R}}
\newcommand{\N}{{\mathbb N}}

\newcommand{\C}{{\mathbb C}}
\newcommand{\Z}{{\mathbb Z}}
\renewcommand{\E}{{\mathbb E}}
\renewcommand{\P}{{\mathbb P}}
\newcommand{\I}{{\mathbb I}}
\newcommand{\T}{{\mathbb T}}

\begin{document}
\title{
Random Unitary Models  and \\ their  Localization Properties}
 \author{ Alain Joye\footnote{ Institut Fourier, UMR 5582,
CNRS-Universit\'e Grenoble I, BP 74, 38402 Saint-Martin
d'H\`eres, France.} \footnote{Partially supported by the Agence Nationale de la Recherche, grant ANR-09-BLAN-0098-01}}

\date{ }

\maketitle
\vspace{-1cm}

\thispagestyle{empty}
\setcounter{page}{1}
\setcounter{section}{1}

\setcounter{section}{0}

\section{Introduction}

This paper aims at presenting a few models of quantum dynamics whose description involves the analysis of random unitary matrices for which dynamical localization has been proven to hold. Some models come from physical approximations leading to effective descriptions of the dynamics of certain random systems that are popular in condensed matter theoretical physics, whereas others find their roots in more abstract considerations and generalizations. Although they may differ in detail, the operators describing the models all have in common the following key features on which their analysis relies heavily: their dynamics is generated by unitary operators on an infinite dimensional underlying Hilbert space which have a band structure when expressed as matrices in a certain basis and the randomness of the models lies in phases of the matrix elements. 

The focus of this note is put on the description of the models and of the localization results available for them. The methods and tools at work in the detailed proofs of these results are only briefly presented, with an emphasis on the similarity with the methods used in the self-adjoint case. A detailed account of such proofs can be found in the paper \cite{HJS} to which the reader is referred for more about technical issues.

The paper starts with a model of electronic dynamics that we call the magnetic ring model and which, in a certain sense, is the root of the other models that follow. The next section makes the connection between the evolution operator of the magnetic ring model and the CMV matrices, which play a major role in the theory of orthogonal polynomials with respect to a measure on the unit circle. Then we introduce the unitary Anderson models as natural $d$-dimensional generalizations based on the structure of the evolution operator stemming from the magnetic ring model, and on its similarity with the well known discrete Anderson model. A final section is devoted to a model of one-dimensional quantum walk in a random environment, another rather popular topic of study in theoretical physics and computer science. For all these models, we state dynamical localization results which are based on the methods that we describe in the last section section of this paper.

\section{Magnetic Ring Model}

Consider an electron in a metallic ring threaded by a time dependent magnetic flux at the center of the ring. Further assume the flux grows linearly with time. According to Maxwell's laws, the flux induces a constant electric force tangent to the ring. Hence the electron is submitted to the field force induced by the periodic metallic background plus the constant force induced by the magnetic flux. A natural question addressed in \cite{lv, bb, ao} consists in asking whether, asymptotically in time, the electron will acquire an unbounded energy due to the constant force it feels or if the defects of the metallic structure of the ring can prevent the growth in energy. 

\begin{figure}[hbt]
\centerline {
\includegraphics[width=8cm]{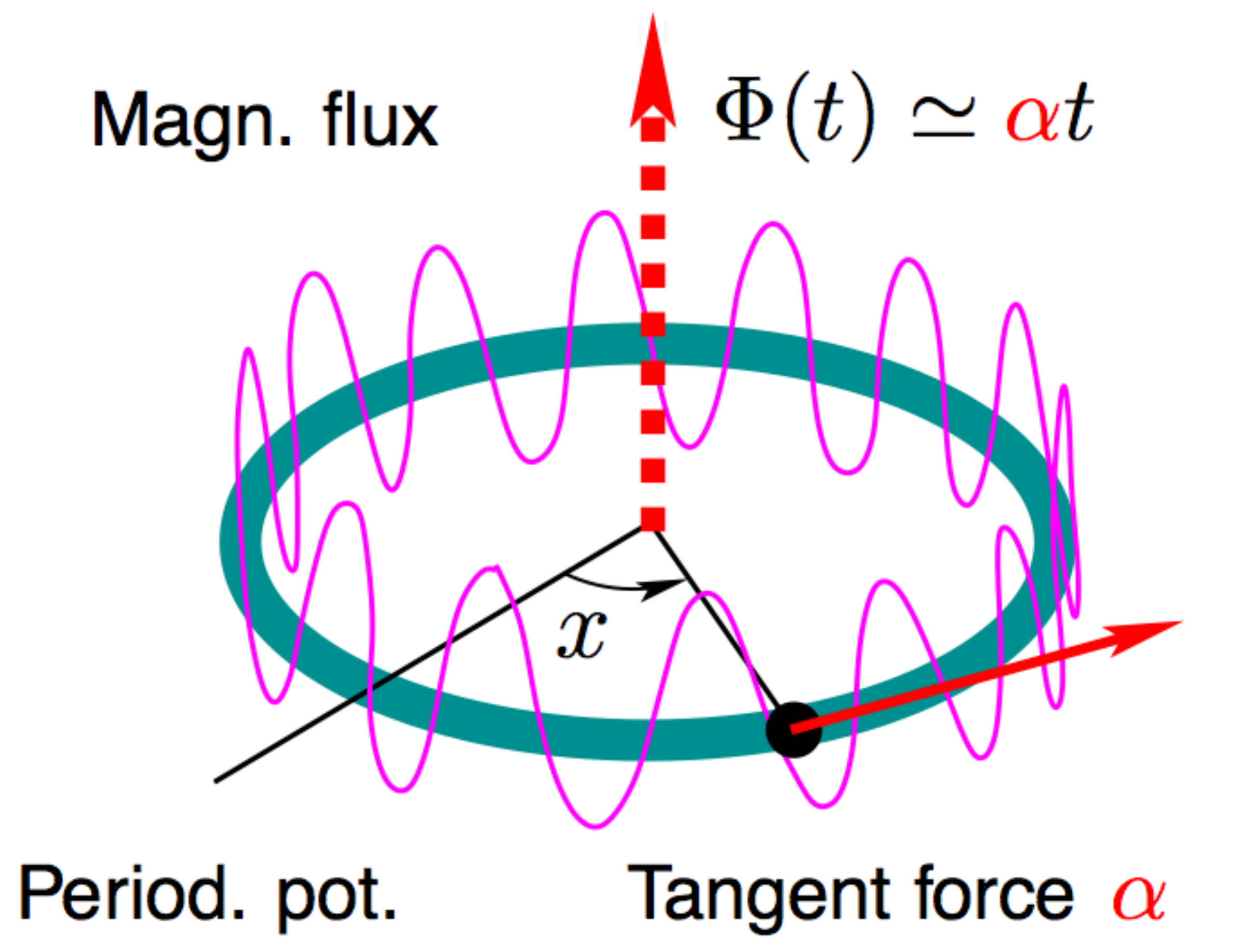}
}
\caption{The magnetic ring model}
\label{ring}
\end{figure}

\begin{figure}[hbt]
\centerline {
\includegraphics[width=8cm]{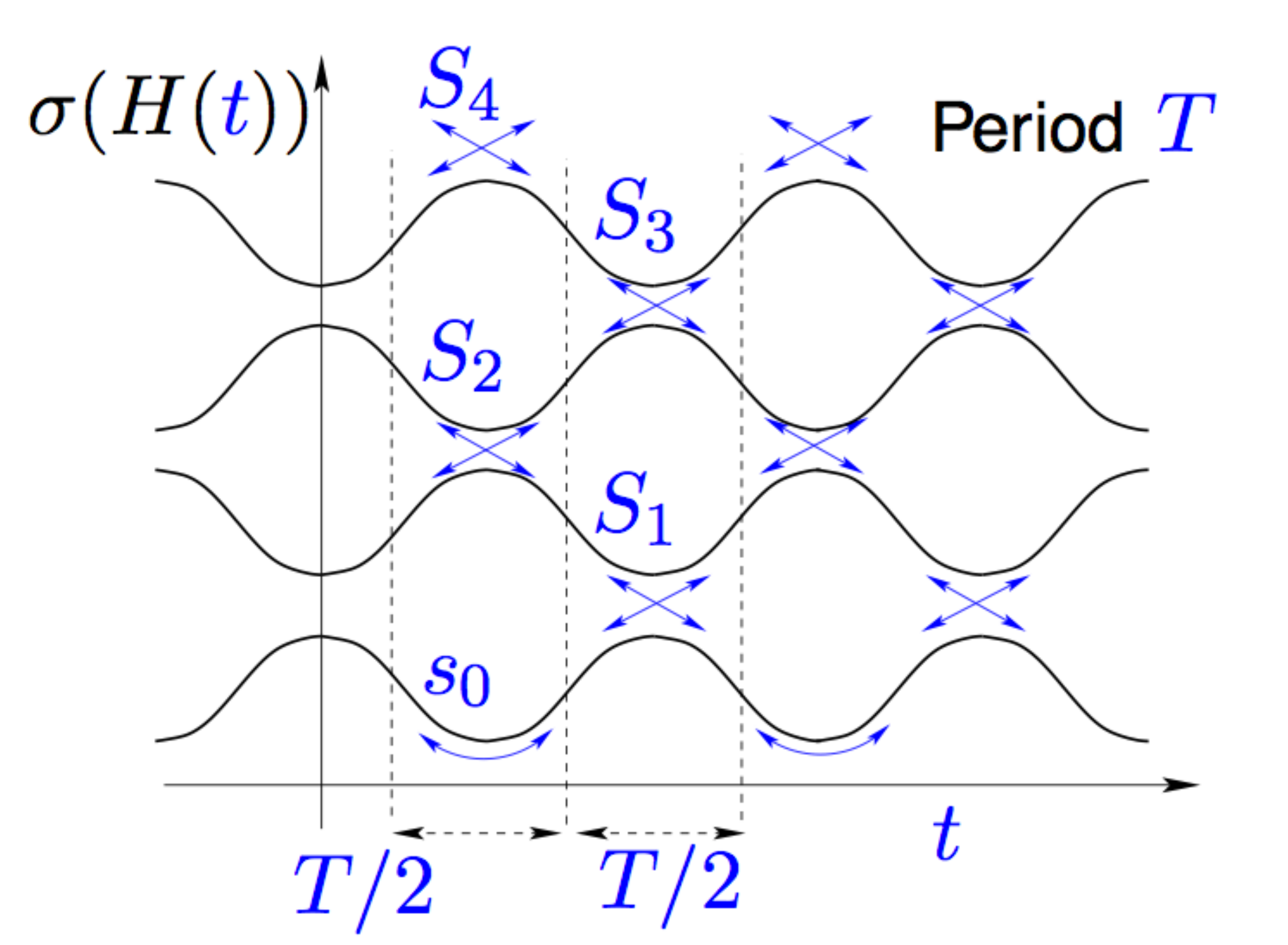}
}
\caption{The energy levels of $H(t)$ and the different transitions considered}
\label{fig1}
\end{figure}

In order to tackle the problem, the following  approximations and regime are considered: the curvature and width of the ring are neglected and the strength of the constant force is small. This leads to an effective one dimensional periodic model in the angular variable, $x\in [0,2\pi) $, see figure \ref{ring}.
The corresponding Hamiltonian takes the form 
\be
H(t)=(-i\partial_x- \alpha t)^2+V_p(x), \ \ \mbox{on }\ \ L^2((0,2\pi]), 
\ee
with periodic boundary conditions, where the parameter $\alpha$ is assumed to be small  and $V_p$ is real valued. Note that the variable $\alpha t$ plays the role of the quasi-momentum for the periodic Schr\"odinger  operator with potential $V_p$ extended to $\R$ by periodicity. Therefore the spectrum of $H(t)$ is given by the corresponding band functions $\{ E_k(t)\}_{k\in\N}$,  and is periodic in $t$. Moreover, the effective Hamiltonian being slowly varying in time for $\alpha<\hspace{-.15cm}<1$,  the adiabatic theorem of quantum mechanics states that an initial condition proportional to an eigenstate of $H(0)$ will give rise at any later time to a solution which belongs to the corresponding eigenspace of $H(t)$ obtained by continuity in time, to leading order in $\alpha$, provided the eigenvalues $E_k(t)$ are simple for all $t$. Therefore, over a period, such an initial state only changes by a phase which depends on the potential $V_p$. In order to describe energy growth, it is necessary to allow transitions between the (instantaneous) energy levels of the model. For a quantitative approach, one makes use of the Landau-Zener formula which says that the amplitude of non adiabatic transitions between two levels is appreciable only when the gap between the levels is small, actually of order $\sqrt{\alpha}$, so that the levels experience an avoided crossing. Now, considering that typically over one period in $t$ each level becomes close to the level immediately above and immediately below only once and at different times, (except for the ground state), see figure \ref{fig1}, 
the effective evolution operator is constructed on the basis of the considerations above as follows: over the first half period, the two levels with indices $2k$ and $2k+1$, $k\geq 0$, exhibit one avoided crossing during that time span and evolve independently of the others, according to some scattering process. Over the next half period, the same scenario takes place, except that the set of independent levels involved in an avoided crossing carry indices $2k-1$, $2k$, (except for the ground state). For a given set of two levels exhibiting an avoided crossing, with indices $k-1$, $k$, with $k\geq 1$,  the scattering process is encoded in a general $2\times2$ unitary matrix  
\bea\label{gen22}
&&S_k=e^{-i\theta_k}
\begin{pmatrix}{r_k e^{-i\alpha_k} & it_k e^{i\gamma_k}\cr
                        it_k e^{-i\gamma_k} & r_k e^{i\alpha_k}
                        }\end{pmatrix}, \ \ \ \mbox{with $ \alpha_k, \gamma_k, \theta_k \in [0,2\pi),$ } \\ \nonumber
&&\hspace{5.8cm}\mbox{ and $r_k,t_k\in [0,1]$, s.t. $r_k^2+t_k^2=1.$}
\eea
The coefficient $t_k$ gives the Landau-Zener transition amplitude associated with the avoided crossing and depends only on the minimum gap displayed by the band functions and their local behavior there. The phases depend in a more complicated way on the global behaviour of the band functions. When $k=0$, $S_0$ is replaced by a phase, $s_0$. In principle, once $V_p$ is given, all parameters of $S_k$ can be computed, within the framework and approximations adopted, see \cite{lv, bb, ao}. Altogether, the effective evolution operator over one period, also called monodromy operator, takes the following matrix form on $l^2(\mathbb N)$ in an orthonormal basis of eigenstates of $H(0)$
\be\label{22blocks}
U=U_o U_e, \ \ \ \mbox{where}\ \ \ 
                  U_o = \begin{pmatrix}{   S_1 \cr
                   &S_{3}& &    \cr
                   & &S_5&   \cr  
                  & & &  \ddots     }\end{pmatrix},
                  U_e = \begin{pmatrix}{  s_0 \cr
                   &S_{2}& &    \cr
                   & &S_4&   \cr 
                  & & &  \ddots  }\end{pmatrix}.
\ee
We shall denote by $\{e_k\}_{k\in \mathbb N}$ the chosen basis such that $H(0)e_k=E_k(0)e_k$, $k\in \N$. 
Note that the $2\times 2$ blocks in  $U_e$ are shifted
by one with respect to those of $U_o$ along the diagonal, and that $s_0$ represents a $1\times 1$ block. Without expliciting the elements, we have the
structure
\be\label{struct}
U=\begin{pmatrix}{
            \ast &\ast &\ast & & & & & &\cr
           \ast &\ast &\ast & & & & & &\cr
            &\ast & \ast&\ast &\ast & & & &\cr
            &\ast &\ast &\ast &\ast & & & &\cr
            & & &\ast &\ast &\ast &\ast & &\cr
            & & &\ast &\ast &\ast &\ast & &\cr 
            & & & & &\ast &\ast &\ast &\ast \cr
            & & & & &\ast &\ast &\ast&\ast  \cr
            & & & & & & & &  \ddots          }\end{pmatrix}   .
\ee
Let us note here that not all phases appearing in the matrix $U$ play a significant role. Indeed, it is shown in \cite{BHJ} that a suitable change of phases of the basis vectors amounts to setting all phases $\{\gamma_k\}_{k\in \N}$ to zero.

On the basis of the arguments leading to the operator $U$ describing the evolution over one period, the large time behavior of the electrons in the ring threaded by a linear magnetic flux is encoded in the properties of the discrete dynamics generated by $U$. This is the starting point of the analysis and we shall not attempt to justify rigorously any of the arguments outlined above.

Coming back to the original motivation, we assume that the periodic potential $V_p$ contains a random component due to the impurities in the metallic ring. Then all matrices $S_k$ are random and, in turn, the monodromy operator becomes a {\it random unitary operator with band structure} that we denote by $U_\omega$. The subscript $\omega$ indicates some configuration of the random parameters. We will specify below the way the monodromy operator depends on the randomness.
Since we are working in an energy eigenbasis $\{e_k\}_{k \in \N}$, the question asked at the beginning of this section can be cast into the following form. 

\bigskip 
{\bf Question:}
Let $\ffi\in l^2(\mathbb N)$ be normalized with compact support, {\it i.e.} $\bra e_k | \ffi \ket =0$, if $k\geq R$, for some $R>0$, so that its energy is bounded above by  $E_R(0)$. For a typical configuration of impurities $\omega$, does the random vector at time $n$ , $U_\omega^n\ffi$, travel to high energy states or spread significantly over high energy states of the basis $e_k$, $k\geq 0$ as $n\ra \infty$ ?  Or does the vector $U_\omega^n\ffi$ remain close to a finite dimensional subspace spanned by basis vectors $e_k$ with $k\leq \rho$, uniformly in $n$ ? A related but not equivalent question is: for a typical configuration $\omega$, does the spectrum of the operator $U_\omega$ contain a continuous component or is it pure point?

\bigskip
We will be able to provide a quantitative answer this question, for certain choices of deterministic and random parameters in the model. We shall refrain from stating results in their full generality, referring the interested reader to the references provided for more details. Several such choices are studied in \cite{ade, BHJ, j, hjs, rhk, dOS}... We will only discuss one of them which, on the one hand, is rich enough for our purpose, and, on the other hand, was actually proposed to study the physical model \cite{lv, bb, ao}. This model is defined as follows:

\medskip

We assume the transition {\it amplitudes} between neighboring levels are deterministic and all take the same value, whereas the phases of the scattering matrices are {\it random}. This hypothesis is certainly a simplification but it also makes the problem more interesting, in the sense that transitions to higher and lower energy levels are equally probable, independently of the energy. Therefore the random phases through their interferences play the key role. See \cite{ade, BHJ} for discussion of cases with variable transition amplitudes. \\

{\bf Assumption A:}\\ 
The coefficents $(t_k,r_r)$ in (\ref{gen22}) all take the same value $(t,r) \in (0,1)^2$, for all $k\geq 0$. \\

We also exclude the trivial case $t=0$ such that $U_\omega$ is diagonal, and $r=0$ such that the absolutely continuous spectrum of $U_\omega$ coincides with the unit circle ${\mathbb S}$, see  Remark \ref{R1} below and \cite{BHJ}.

\medskip

Next, we assume the randomness enters the operator $U_\omega$ through phases which are i.i.d. on the unit circle. 
We formalize this as follows. Let  $(\Omega, {\cal F}, \P)$ be a probability space, where
$\Omega$ is identified with $\{{\T}^{\N} \}$, $\T = \R / 2\pi\Z$
being the torus, and $\P=\otimes_{k\in\N}\P_k$, where $\P_{k}=\nu$
for any $k\in\N$ and $\nu$ is a fixed probability measure on $\T$,
and ${\cal F}$ the $\sigma$-algebra generated by the cylinders. We
define a set of random variables on $(\Omega, {\cal F}, \P)$  by
\bea\label{beta} \theta_k: \Omega \rightarrow \T,
  \ \ \mbox{s.t.} \ \ \theta_k^\omega=\omega_{k}, \ \ \ k\in \N.
\eea These random variables $\{\theta_k\}_{k\in\N}$ are thus i.i.d.\
on $\T$.

\medskip
{\bf Assumption B:}\\
Let $D_{\omega}=\mbox{ diag }\{e^{-i\theta_k^\omega}\}$ in the basis $\{e_k\}_{k\in\N}$, where the $\theta_k^\omega$'s are given in (\ref{beta}). Suppose
$d\nu(\tau)=\tau(\theta)d\theta $, where $0\leq \tau \in L^\infty([0,2\pi))$. 
\medskip

Under Assumptions A and B, we consider operators $U_\omega$ of the form
\be\label{1d0}
 U_{\omega}=D_{\omega}S, \,\,\,\mbox{ with }
D_{\omega}=\mbox{ diag }\{e^{-i\theta_k^\omega}\}
\ee
and
\be\label{s0}
S=\begin{pmatrix}{
               r& rt & -t^2& & & \cr
              -t & r^2& -rt  & & & \cr
               & rt & r^2 & rt & -t^2& \cr
               & -t^2 &-tr & r^2& -rt& \cr
               & & & rt &r^2 & \cr
               & & & -t^2& -tr&\ddots }\end{pmatrix}.
\ee 

In the case where all the (relevant) phases in the scattering matrices $S_k$ are i.i.d. and uniform on the unit circle, it can be shown that  $U_\omega$ takes the form (\ref{1d0}) with a uniform density $\tau$, see \cite{BHJ}. This special case is argued to be physically relevant in \cite{bb}, but the result below holds for any density $\tau$ satisfying assumption B. Note that the operator $S$ is obtained by formula (\ref{22blocks}) with blocks $S_k$ of the form
\be
S_{2k+1}=\begin{pmatrix}{r& t\cr
                        -t & r
                        }\end{pmatrix} , \ \ \ S_{2(k+1)}=\begin{pmatrix}{r& -t\cr
                        t & r
                        }\end{pmatrix}, \ \ \ \forall k\in \N, \ \ \mbox{and} \ \  s_0=1.
\ee\\

\begin{thm}\label{T1} \cite{HJS} Consider $U_\omega$ defined in (\ref{22blocks}), under  assumptions A and B. Let $t\in (0,1)$ be arbitrary and denote by $\E$ the expectation over $\omega$. Then there exist $\alpha>0$, $C<\infty$ such that 
\be \label{espun}
\E\left[\sup_{n\in \Z}\left|\bra e_j | U_\omega^n e_k\ket\right|\right]\leq Ce^{-\alpha |j-k|}.
\ee
Consequently, for any $p>0$, we have
\be \label{mom}
\sup_{n\in\Z}\|X^pU_\omega^n\ffi\|^2 < \infty \ \ \mbox{almost surely,}
\ee
where the operator $X$ is defined by $X e_k=k e_k$, for all $k\in \N$. Moreover, the spectrum of $U_\omega$ is pure point: 
\be
\sigma(U_\omega)=\sigma_{pp}(U_\omega)\ \ \mbox{almost surely}
\ee
with exponentially decaying eigenfunctions.
\end{thm}

The previous statement is a {\it dynamical localization result} in energy space. Further assuming that $E_k(0)\leq C' k^p$, as $k\ra \infty$ for some $C', p <\infty$, it shows that  the energy of the electron in the disordered metallic ring does not grow unboundedly with time, despite the constant force acting on it. Also, the probability to find the electron in high energy states, {\it i.e.} with high quantum number number, decays faster than any inverse power of the quantum number. Note however, that there are different circumstances where the spectrum of $U$ may be pure point but the energy can grow in time, \cite{dOS}.

\medskip

\begin{rem} \label{R1}
It is often technically simpler to consider that the operator $U_\omega$ acts on $l^2(\Z)$ rather than on $l^2(\N)$. This means that  all indices $k$ are considered as elements of $\Z$ instead of $\N$, that $\Omega=\{\T^{\Z}\}$, $\P=\otimes_{k\in\Z}\nu$, and that we deal with 
unitary operators of the form
\be\label{1d}
 U_{\omega}=D_{\omega}S, \,\,\,\mbox{ with }
D_{\omega}=\mbox{ diag }\{e^{-i\theta_k^\omega}\}
\ee
and
\be\label{s2}
S=\begin{pmatrix}{\ddots & rt & -t^2& & & \cr
              & r^2& -rt  & & & \cr
              & rt & r^2 & rt & -t^2& \cr
              & -t^2 &-tr & r^2& -rt& \cr
              & & & rt &r^2 & \cr
              & & & -t^2& -tr&\ddots }\end{pmatrix}
\ee where the translation along the diagonal is fixed by $\bra
e_{2k-2}|S  e_{2k} \ket =-t^2$, $k\in\Z$. 
\end{rem}

In particular, on $l^2(\Z)$, one sees rightaway that if $r=0$, $U_\omega $ is unitarily equivalent to a direct sum of two shifts.  Hence it has  purely absolutely continuous spectrum given by ${\mathbb S}$. Since one can pass from $U_\omega$ defined on $l^2(\Z)$ to two copies of the monodromy operator defined on $l^2(\N)$ by a finite rank operator, this shows that $\sigma_{a.c.}(U_\omega)= {\mathbb S}$ in either case.

Theorem \ref{T1} applies to this setting as well, {\it mutatis mutandis}, as discussed in \cite{BHJ, hjs}.

\section{Orthogonal Polynomials on the Unit Circle}

Before we turn to other generalizations of this model, we briefly mention in this section that unitary operators with a band structure of the form (\ref{struct}) appear naturally in the theory of orthogonal polynomials on the unit circle. For a detailed account of this topic, we refer to the monograph \cite{s}. Given  an infinitely supported probability measure $d\mu$ on ${\mathbb S}$, such
polynomials $\Phi_k$ are determined via the recursion
\be
\Phi_{k+1}(z) = z\Phi_k(z)-\overline{\alpha}_k \Phi_k^*(z), \ \ \ \mbox{with} \ \ \
\Phi_k^*(z)=z^k\overline{\Phi_k(1/\overline z)}, \ \ 
\Phi_0=1,
\ee by a sequence of complex valued coefficients
$\{\alpha_k\}_{k\in\N}$, such that $|\alpha_k|< 1$, called
Verblunsky coefficients, which also characterize the measure
$d\mu$, see \cite{s}. This latter relation is encoded in a five
diagonal unitary matrix $C$ on $l^2(\N)$ representing
multiplication by $z\in {\mathbb S}$: the measure $d\mu$ arises as the
spectral measure $\mu(\Delta)= \bra e_0|E(\Delta)e_0\ket$ of the
cyclic vector $e_0$ of $C$, where $dE$ denotes the spectral family of $C$. This matrix is the equivalent of
the Jacobi matrix in the case of orthogonal polynomials with
respect to a measure on the real axis, and it is called the CMV
matrix, after \cite{cmv}.

Writing the Verblunsky coefficients as
\be 
\alpha_k=re^{i\eta_k}, \ \ \ \mbox{and setting }  \ \ t_k= \sqrt{1-r_k^2}, \ \ \ k=0,1, \ldots ,
\ee 
the corresponding CMV matrix reads 
\be C=\begin{pmatrix}{  r_0e^{-i\eta_0} & r_1t_0e^{-i\eta_1} & t_0t_1 & & & \cr
               t_0& -r_0r_1e^{i(\eta_0-\eta_1)}  &-r_0t_1e^{i\eta_0} & & &\cr
                & r_2t_1e^{-i\eta_2} & -r_1r_2e^{i(\eta_1-\eta_2)} & r_3t_2e^{-i\eta_3}& t_2t_3 &\cr
                &t_1t_2 & -r_1t_2e^{i\eta_1}& -r_2r_3e^{i(\eta_2-\eta_3)}&-r_2t_3e^{i\eta_2}& \cr
               & & &r_4t_3e^{-i\eta_4} & -r_3r_4e^{i(\eta_3-\eta_4)} & \cr
               & & &t_3t_4 &-r_3t_4e^{i\eta_3} &\ddots
               }\end{pmatrix}
\ee 
which is a special case of (\ref{22blocks}), see {\it e.g.} \cite{j}. In the same way as tri-diagonal Jacobi matrices can be seen as paradigms for self-adjoint operators, the result of \cite{cmv} shows that five-diagonal unitary matrices (\ref{22blocks}) are paradigms of unitary operators. This gives a model independent motivation for the study of such operators.

Comparing with (\ref{1d0}), it was noted in \cite{hjs} that if  the Verblunsky coefficients all have the same modulus and if their phases $\eta_k=\theta_k+\theta_{k-1}+\cdots+\theta_0$, then $C$ is unitarily equivalent to $-U$. Therefore, assuming the $\theta_k^\omega$ are i.i.d., Theorem 1 then directly yields the
\begin{cor}  \cite{hjs, HJS}\\
Let ${\alpha_k(\omega)}_{k\in\N_0}$ be random Verblunsky
coefficients of the form \be\alpha_k(\omega)=re^{i\eta_k(\omega)},
\ \ \  0<r<1, \ \ \ k=0,1,2,\ldots \ee whose phases are
distributed on $\T$ according to \be \eta_k(\omega) \sim d\nu *
d\nu * \cdots
* d\nu\,, \ \ \ \mbox{($k+1$ convolutions)} 
\ee 
where
$d\nu$ satisfies assumption B. Then, the random
measure $d\mu_\omega$ on ${\mathbb S}$ with respect to which the
corresponding random polynomials $\Phi_{k,\omega}$ are orthogonal
is almost surely pure point. Moreover, both (\ref{espun}) and (\ref{mom}) hold.
\end{cor}
\begin{rem} Other dynamical localization results for random polynomials on
the unit circle are proven for
independent Verblunsky coefficients, \cite{ps, t, su}. The results of
\cite{su} and \cite{ps} require rotational invariance of the
distribution of the Verblunsky coefficients in the unit disk. By
contrast, the corollary above holds for strongly correlated
random Verblunsky coefficients.
\end{rem}

\section{Unitary Anderson Models}

When the unitary operator $U_\omega=D_\omega S$ is considered on $l^2(\Z)$ according to Remark \ref{R1}, the similarity with the well known (self-adjoint) one-dimensional discrete Anderson model is evident: The 2-translation invariant unitary operator $S$ given in (\ref{s2}) plays the role of the translation invariant discrete Laplacian $\Delta$ and the diagonal random matrix $D_\omega$ is similar to the diagonal random potential operator $V_\omega$. The sum $-\Delta+V_\omega$ is replaced by the product $D_\omega S$, since we deal with unitary operators. Although $U_\omega\neq e^{-i(\Delta+V_\omega)}$, this operator can be viewed as an effective generator of a discrete dynamics of a particle on the one dimensional lattice. In that case, Theorem \ref{T1} can be interpreted as dynamical localization result in a one dimensional configuration lattice, which begs to be generalized to arbitrary dimension. Such a generalization was proposed in \cite{j2} which we now describe.

\medskip

To define the multidimensional version of the unitary equivalent of the Laplacian, we  view $l^2(\Z^d)$ as $\otimes_{j=1}^d
l^2(\Z)$ and define the canonical basis vectors $e_k$, for $k\in\Z^d$ by $e_k\simeq
e_{k_1}\otimes...\otimes e_{k_d}$. Making explicit the dependence in $t$ in  $S=S(t)$ from (\ref{s2}), we define $S_d(t)$ by
\be
\label{highdimS}
S_d(t) =\otimes_{j=1}^d S(t).
\ee
We denote by $|\cdot|$ the maximum norm on $\Z^d$. Using this norm it is easy to see that $S_d(t)$ inherits
the band structure of $S(t)$ so that
\be
\langle e_k| S_d(t)
e_l\rangle=0 \qquad \mbox{if } |k-l|>2.
\ee
Due to the tensor product structure, the spectrum of $S_d(t)$ is obtained from that of $S(t)$, which can be determined by using Fourier transform. We get
\be
\sigma(S_d(t)) = \{e^{i \vartheta}: \vartheta \in [-d\lambda_0,d\lambda_0]\}, \ \ \ \mbox{where} \ \ \ \lambda_0=\arccos (1-2t^2).
\ee

The random operator $D_\omega$ keeps the same form in the canonical basis, $D_\omega=\mbox{ diag }\{e^{-i\theta_k^\omega}\}$, with the understanding that $\{\theta_k^\omega\}_{k\in\Z^d}$ are i.i.d. on $\T$, with distribution $d\nu$. 

\medskip

The operator  
\be\label{uam}
U_\omega=D_\omega S_d(t) \ \ \ \mbox{defined on $l^2(\Z^d)$}
\ee
is called the generator of the {\it unitary Anderson model}.

\medskip

In that framework, Theorem \ref{T1} is a unitary version of the statement that dynamical localization holds true for any disorder strength in one dimension for the Anderson model with absolutely continuous distribution of potential. 
As is well known, localization results for the Anderson model in two and higher dimensions are only available in certain asymptotic regimes of the parameters, typically large disorder, or in certain subsets of the spectrum, the band edges.  We state below two localization results which hold in the same regimes.
The dynamical localization
property in $\Z^d$ is measured in terms of the boundedness in time of all
quantum moments of the position operator on the lattice. More
precisely, for $p>0$ we let $|X|_e^p$ be the maximal multiplication operator such that
\be
|X|_e^pe_j=|j|_e^p e_j, \ \ \ \mbox{for } j\in{\mathbb Z}^d,
\ee
where $|j|_e$ denotes the Euclidean norm on $\Z^d$.

\medskip

For the unitary Anderson model the parameter $t$ takes the role of a disorder parameter. Small values of $t$ correspond to large disorder in the sense that $U_{\omega}$ is dominated by its diagonal part, since $S_d(t)$ tends to the identity as $t\ra 0$.  
The following result says that in any dimension, dynamical localization holds throughout the spectrum of $U_\omega$, provided $t$ is small enough:
\begin{thm}\label{Tld} \cite{j2, HJS}
Consider $U_\omega$ defined by (\ref{uam}), under assumption B. Then, there exists $t_0>0$ such that for all $t<t_0$, $\sigma(U_\omega)=\sigma_{pp}(U_\omega)$ almost surely. Moreover, there exist $\alpha>0$, $C<\infty$ such that for all $j,k\in \Z^d$
\be 
\E\left[\sup_{n\in \Z}\left|\bra e_j | U_\omega^n e_k\ket\right|\right]\leq Ce^{-\alpha |j-k|}.
\ee
Consequently, for any $p\geq 0$ and for any
$\ffi$ in $l^2({\mathbb Z}^d)$ of compact support,
\be
\sup_{n\in {\mathbb Z}}\||X|_e^p U_\omega^n\ffi\|<\infty \ \ \mbox{almost surely.}
\ee
\end{thm}

\medskip

Let us consider now the band edge regime. 
At this point, it is useful to point out that the periodicity along the diagonal of the matrix  $S$ and the definition of $D_\omega$ make the operator
$U_\omega$ ergodic with
respect to the $2$-shift in $\Omega=\T^{\Z^k}$. By the general
theory of ergodic operators, see
\cite{CL}, it follows that the spectrum
of $U_\omega$ is almost surely deterministic, i.e.\ there is a
subset $\Sigma$ of the unit circle such that $\sigma(U_\omega) =
\Sigma$ for almost every $\omega$. The same is true for the
absolutely continuous, singular continuous and pure point parts of
the spectrum. Explicitely, there are $\Sigma_{ac}$, $\Sigma_{sc}$ and
$\Sigma_{pp}$ such that almost surely $\sigma_{ac}(U_{\omega}) =
\Sigma_{ac}$, $\sigma_{sc}(U_\omega) = \Sigma_{sc}$ and
$\sigma_{pp}(U_\omega) = \Sigma_{pp}$. Moreover, $\Sigma$ can be characterized in terms of the support of
$\nu$ and of the spectrum of $S_d(t)$, \cite{j}:
 \be \label{eq:asspectrum}
\Sigma =\exp{(-i\,\mbox{supp}\,\nu)}\,\sigma(S_d(t))=
\{e^{i\alpha}:\alpha\in [-d\lambda_0,d\lambda_0]- \mbox{supp}\,\nu \}.
\ee
These facts also hold for the one dimensional half lattice operator (\ref{1d0}).

For simplicity, and without loss of generality, we assume that supp$\,\nu \subset [-\beta,\beta]$ with $\beta\in (0,\pi)$ and $-\beta, \beta\in \mbox{supp}\,\nu$.
Furthermore, we will work under

{\bf Assumption C:}
\be
\beta + d\lambda_0 < \pi.
\ee

\begin{figure}[hbt]
\centerline {
\includegraphics[width=11cm]{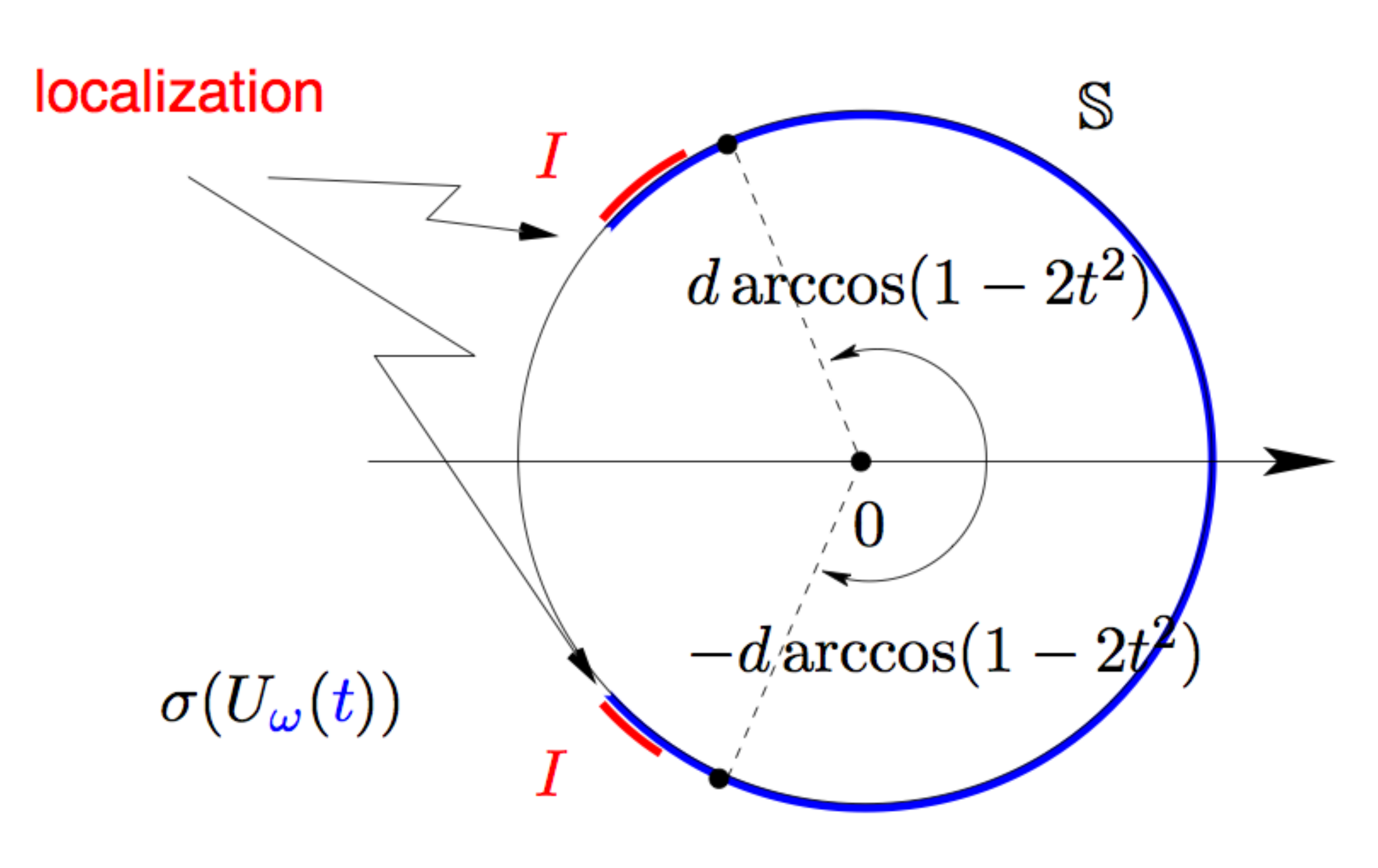}
}
\caption{The spectrum of $U_\omega$ and regions of localization $I$.}
\label{edge}
\end{figure}

By (\ref{eq:asspectrum}), this implies the existence of a gap in the almost sure spectrum $\Sigma$ of $U_{\omega}$,
\be
\{e^{i\vartheta}: \, \vartheta \in (d\lambda_0+\beta, 2\pi -d\lambda_0-\beta)\} \cap \Sigma = \emptyset,
\ee
and that $e^{i(d\lambda_0+\beta)}$ and $e^{i(2\pi -d\lambda_0-\beta)}$ are band edges of $\Sigma$. In any dimension, and for any disorder, the result below states that localization takes place  at the band edges, at arcs denotes by $I$ in figure \ref{edge}.

To focus on specific parts of the spectrum of $U_\omega$, we introduce spectral projectors $P^\omega_{[a,b]} $ on intervals $[a,b]\subset \T$, by 
$P^\omega_{[a,b]} =E^\omega([e^{ia}, e^{ib}])$, where $dE^\omega$ is the spectral family of $U_\omega$ and $[e^{ia}, e^{ib}]$ is a positively oriented arc on $ {\mathbb S}$.

\begin{thm}\label{dynspec} Consider $U_\omega$ defined by (\ref{uam}), under assumptions B and C. Then, there exists $\gamma >0$ such that for the interval $[a,b]=[d\lambda_0+\beta-\gamma, d\lambda_0+\beta]$  it holds:
\be
(a,b)\cap\Sigma\neq \emptyset\ \ \ \mbox{and}\ \ \ (a,b)\cap\Sigma_{cont}=\emptyset,
\ee
where $\Sigma_{cont}=\Sigma_{ sc} \cup \Sigma_{ac}$. In other words, almost surely
$P_{[a,b]}^\omega U_\omega$ has pure point spectrum.
Moreover, there exist constants $C<\infty$ and $\alpha>0$ such that
\be
 \E[\sup_{n\in\Z} |\langle e_j|
U_{\omega}^n P^\omega_{[a,b]} e_k \rangle|] \le Ce^{-\alpha
|j-k|} \ee for all $j,k \in \Z^d$. 
And, consequently, for any $p\geq 0$ and for any
$\ffi$ in $l^2({\mathbb Z}^d)$ with compact support,
\be
\sup_{n\in {\mathbb Z}}\||X|_e^p U_\omega^nP_{[a,b]}^\omega\ffi\|<\infty \ \ \mbox{almost surely.}
\ee
The same is true for the interval $[a,b]=[2\pi -d\lambda_0-\beta, 2\pi -d\lambda_0-\beta+\gamma]$.

\end{thm}

\section{Quantum Walks in Random Environments}

We now turn to another type of discrete quantum dynamics which can be shown to display localization due to the presence of disorder. Quantum walks have become a popular topic of study due to the role they play in theoretical quantum computing, see {\it e.g.} \cite{M}, \cite{Ke}, \cite{SKW}, \cite{S},..., to their use in the description of effective dynamics of quantum systems, see \cite{ADZ}, \cite{Ketal}, \cite{Zetal}, and to the nice mathematical properties they have, \cite{AAKV}, \cite{Ko}, \cite{CGMV}.

\medskip

Let us consider the simplest instance of a quantum walk, {\it i.e.} a quantum walk on $\Z$. Such walks simply describe the discrete dynamics of a quantum particle with spin. In this context, the spin state is often called {\it coin} state. The Hilbert space is thus  
\begin{equation}
\ch = \cx^2\otimes l^2(\cz).
\label{m1}
\end{equation}
We denote by $\{\upket, \downket\}$ a canonical basis of $\cx^2$ and we denote the (position) canonical basis vectors of $l^2(\Z)$ by $\{|n\rangle\}$, $n\in\cz$.
The time-one dynamics of the system is composed of two steps: a unitary evolution of the spin alone by means of a unitary operator $C$ on $\cx^2$, followed by the motion of the walker, conditioned on the spin state. More precisely, if the spin is pointing up the walker moves to the right one step, and if the spin is pointing down the walker moves to the left. The latter step is determined by the action implemented by the unitary operator
\begin{equation}
S=\sum_{k\in \cz} \left\{ P_\uparrow\otimes|k+1\rangle\langle k| +  P_\downarrow\otimes|k-1\rangle\langle k|\right\}
\end{equation}
where we have introduced the orthogonal projections
\begin{equation}
P_\uparrow = \upket\upbra \mbox{\quad and \quad} P_\downarrow = \downket\downbra .
\label{m4}
\end{equation}
Altogether, the one step dynamics consists in shuffling the spin variable and then performing the spin dependent shift $S$:
\begin{equation}
\label{a1}
U=S(C\otimes {\mathbb I}) \quad \quad  \mbox{with}\quad \quad 
C=\left[\begin{array}{cc}
a& b\\ c&d
\end{array}\right] \quad \quad \mbox{s.t. } \quad \quad C^*=C^{-1}.
\end{equation}
The evolution operator at time  $n$ then reads $U^n$.

\medskip

Hence, if one starts form the state $\upket\otimes|k\rangle$, the (quantum) probability to reach, in one time step, the site $|k+1\rangle$ equals $|a|^2$ whereas that to reach $|k-1\rangle$ equals $1-|a|^2$. Similarly, starting from $\downket\otimes|k\rangle$, the probability to reach the site $|k+1\rangle$ equals $|b|^2$ and that to reach $|k-1\rangle$ is $1-|b|^2$. The similarity in interpretation with a classical random walk explains why the spin variable and the operator $C$ are  called the coin states and coin operator. Despite the similarity of this dynamics with that of a classical random walk, there is nothing random in the quantum dynamical system at hand. The dynamics is invariant under translations on the lattice $\Z$, which implies ballistic transport.

\medskip

More quantitatively, 
let  $X=\un\otimes x$ denote the operator defined on (its maximal domain in) $\cx^2\otimes l^2(\cz)$, where $x$ is the position operator given by $x |k\ket=k|k\ket$, for all $k\in\Z$. For any $p>0$, $n\in \Z$, and any $\ffi$ with compact support, we consider the quantum mechanical expectation of the operator $X$ at time $n$ by
\begin{equation}\label{defxx}
\bra  X^p \ket_{\ffi}(n)= \bra \ffi | U^{-n} X^{p} U^{n} \ffi\ket = \|X^{p/2} U^{n} \ffi\|^2.
\end{equation}
The analog definition holds for $\bra  |X|^p \ket_{\ffi}(n)$. By Fourier transform methods, one gets

\begin{lem}
 \label{ballistic} For any $\ffi\in \ch$ with compact support,
$$
\lim_{n\ra\infty}\frac{\bra X^2\ket_{\Psi}(n)}{n^2}=B\geq 0
$$ 
with $B=0$ iff $C$ is off diagonal.
\end{lem}
When $C$ is off diagonal, complete localization takes place.

A quantum walk in a non-trivial environment is characterized by coin operators that depends on the position of the walker: for every $k\in\cz$ we have a unitary $C_k$ on $\cx^2$, and the one step dynamics is given by
\begin{equation}
U = \sum_{k\in \cz} \left\{ P_\uparrow C_k\otimes|k+1\rangle\langle k| + P_\downarrow C_k\otimes|k-1\rangle\langle k|\right\}.
\label{m5}
\end{equation}
We consider a {\it random environment} in which the coin operator $C_k$ is a {\it random} element of $U(2)$, satisfying the following requirements:
\medskip

\noindent
{\bf Assumption D:}

\noindent
{\bf (a)} $\{ C_k\}_{k\in\Z}$ are independent and identically distributed $U(2)$-valued random variables.

\noindent
{\bf (b)} The quantum {\it amplitudes} of the transitions to the right and to the left are independent random variables.

\noindent
{\bf (c)} The quantum transition {\it probabilities} between neighbouring sites are deterministic and  independent of the site.   

\medskip

There are of course plenty of meaningful ways to define a (random) environment for a quantum walk, see {\it e.g. }\cite{SBBH, KLMW, YKE, Ko1, sk}. Assumption D is motivated by simplicity and by proximity with the classical random walk. It turns out this choice actually dictates the form of the random coin operators as follows.

\begin{lem}\label{invC}\cite{jm}
Under  Assumption D, the operator $U_\omega$ defined by (\ref{m5}) is unitarily equivalent to the one defined by the choice
\begin{equation}\label{defcoin}
 \left[
\begin{array}{cc}
e^{-\i\omega_k^\uparrow} t & -e^{-\i\omega_k^\uparrow} r\\
e^{-\i\omega_k^\downarrow} r & e^{-\i\omega_k^\downarrow} t
\end{array}
\right] \ \ \ \ \ \ \mbox{\quad where $0\leq t,r\leq 1$ and $r^2+t^2=1$}
\end{equation}
and  $\{\omega_k^\uparrow\}_{k\in {\mathbb Z}}\cup\{\omega_k^\downarrow\}_{k\in{\mathbb Z}}$ are i.i.d. random variables defined as in  (\ref{beta}), 
up to multiplication by a global deterministic phase.
\end{lem}

Let $U_\omega$ be the one step dynamics of a quantum walk in a random environment defined by (\ref{m5}) with $C_k$, $k\in\Z$ given by (\ref{defcoin}), where 
$\{\omega_k^\#\}_{k\in \Z, \#\in\{\uparrow, \downarrow\} }$ are the i.i.d. random variables  defined in (\ref{beta}), distributed according to an absolutely continuous measure $\nu$ on $\T$. Then a statement equivalent to Theorem \ref{T1} in this context holds.

\begin{thm} \label{QW} \cite{jm} Assume B holds for the distribution $d\nu$. Then, for any $t\in (0,1)$, $$\sigma(U_\omega)=\sigma_{pp}(U_\omega) \ \mbox{almost surely}.$$ 
Moreover, there exist $C<\infty$, $\alpha>0$ such that for any $j, k \in \Z$ and any $\sigma, \tau \in \{\uparrow, \downarrow\}$
\begin{equation}\label{loces}
\E \left[ {\sup_{n\in \Z}}\  |\bra \sigma \otimes j | U_\omega^n \, \tau \otimes k \ket | \right]\leq Ce^{-\alpha |j-k|}
\end{equation}
and, for any $p>0$, almost surely, 
\be
\sup_{n\in\Z} \bra  X^p \ket^\omega_{\ffi}(n) < \infty.
\ee
\end{thm} 

\medskip

The similarity in this result and Theorem \ref{T1} stems from the similarity of the random unitary operators in the two cases considered. More specifically, Lemma \ref{invC} shows that, up to unitary equivalence and multiplication by a global phase,  $U_\omega$ has the following representation in the 
ordered basis $\{e_k\}_{k\in\Z}=\{\ldots,\upket\otimes|n-1\rangle, \downket\otimes|n-1\rangle, \upket\otimes|n\rangle, \downket\otimes|n\rangle,\ldots\}$, 
\begin{equation}
U_\omega= D_\omega S,\mbox{\quad with\quad} S=
\left[
\begin{array}{cccccccc}
\ddots & r & t &   &   & &  &  \\
 & 0 & 0 &  &  &   &  &\\
 & 0 & 0 & r & t  &   &  &\\
 & t  & -r & 0 & 0 &   &  &\\
 &   &   & 0 & 0 & r & t &\\
 &   &   & t  & -r & 0 & 0&\\\vspace*{-2.5mm}
 &   &   &   &   & 0 & 0& {}_{\ddots}\\
 &   &   &   &   & t  & -r & 
\end{array}
\right].
\label{71}
\end{equation}
Here the diagonal of $S$ consists of zeroes and the labeling of the basis is such that the odd rows contain $r, t$ and the even rows contain $ t, -r$. Moreover, upon relabeling the indices of the random phases, $D_\omega$ is diagonal with i.i.d. entries, $D_\omega={\rm diag}(\ldots,e^{-\i\theta^\omega_k}, e^{-\i\theta^\omega_{k+1}},\ldots)$.  

Note that since the random operator at hand differs from that of Remark \ref{R1} by the form of the deterministic matrix $S$, the   localization result stated in Theorem \ref{QW} requires the separate analysis provided in \cite{jm}.

\section{Methods}

Now that we have described several similar random unitary operators appearing in the study of different quantum models, we want to address the methods used to derive dynamical localization results for these operators. The paper \cite{HJS} is devoted to a detailed and hopefully pedagogical exposition of these methods, so we only point out here the main steps of the analysis. As mentioned already, the analysis draws on the similarity of these random unitary models with the self-adjoint discrete Anderson model. Actually, our approach to localization proofs will be via a unitary version of the fractional moment method, which was initiated as a tool in the theory of selfadjoint Anderson models by Aizenman and Molchanov in \cite{AM}. Dynamical localization will follow as a general consequence of exponential decay of spatial correlations in the fractional moments of Green's function. 

\medskip

Let us consider a random unitary matrix with a band structure in a distinguished basis $\{e_k\}_{k\in\Z^k}$ of $l^2(\Z^d)$ of the form
\be
 U_{\omega}=D_{\omega}S_d, \,\,\,\mbox{ with }
D_{\omega}=\mbox{ diag }\{e^{-i\theta^\omega_k}\}
\ee
where the random phases $\{e^{-i\theta^\omega_k}\}_{k\in\Z^d}$ satisfy assumption B (adapted to the $d$-dimensional setting) and the matrix $S_d$ is a $d$-fold tensor product of the five-diagonal unitary operators (\ref{s2}) invariant under the 2-shift. Again, some results hold under weaker hypotheses, but we stick to our setting in order to keep things simple.

Let 
\be
G_\omega(k,l;z)=\bra e_k | (U_\omega -z)^{-1} e_l\ket 
\ee
be the Green function of $U_\omega$ defined for $z\in \rho(U_\omega)$, the resolvent set of $U_\omega$. Now, the structure of $U_\omega$ is such that a modification in one of the random parameters corresponds to a rank one perturbation of the original operator. This leads to the observation that while the Green function becomes singular as $z$ approaches the spectrum of $U_{\omega}$, these singularities are fractionally integrable with respect to the random parameters: for $s\in (0,1)$ the {\it fractional moments of the resolvent}, $\E(|G(k,l;z)|^s)$, have bounds which are uniform for $z$ arbitrarily close to the spectrum. This is the content of our first result.

\begin{thm} \label{thm:fmbound}
Suppose assumption B holds for the random variables $\{\theta_k\}_{k\in \Z^d}$.
Then for every $s\in (0,1)$ there exists $C(s) <\infty$ such that
\be \label{eq:fmbound}
\int\int |G_\omega(k,l;z)|^s d\nu(\theta_k) d\nu(\theta_l) \le C(s)
\ee
for all $z\in
\C$, $|z|\not=1$, all $k, l \in \Z^d$, and arbitrary values of $\theta_j$, $j\not\in \{k,l\}$.
Consequently,
\be\label{eq:fmbound2}
\E(| G_\omega(k,l;z)|^s) \le C(s),
\ee
for all $z\in
\C$, $|z|\not=1$.
\end{thm}
\begin{rem}
The proof of this general result makes use of the fact that the measure $d\nu$ has a density in $L^\infty$ .
\end{rem}

Then, the goal is to make use of the specificities of the model under study to identify regimes or situations where the fractional moments $\E(|G(k,l;z)|^s)$ are not just uniformly bounded, but {\it decay exponentially} in the distance between $k$ and $l$. The following general result shows that this can be used as a criterion for dynamical
localization of $U_{\omega}$.

\begin{thm} \label{thm:dynamicallocalization} 
Suppose assumption B holds for the random variables $\{\theta_k\}_{k\in \Z^d}$
and that for some $s\in (0,1)$, $C<\infty$,
$\alpha>0$, $\varepsilon >0$ and an interval $[a,b]\in \T$,
\be \label{eq:fmexpdecay}
\E(|G(k,l;z)|^s) \le Ce^{-\alpha|k-l|}
\ee
for all $k,l \in \Z^d$ and all $z\in \C$ such that $1-\varepsilon <
|z|<1$ and arg$\,z \in [a,b]$.

Then there exists $\tilde{C}$ such that \be
\label{eq:dynamicallocalization} \E[\sup_{ {f\in C({\mathbb
S})}\atop {\|f\|_{\infty}\le 1}} |\langle e_k|
f(U_{\omega}) P_{[a,b]}^{\omega} e_l \rangle|] \le \tilde{C}e^{-\alpha
|k-l|/4} \ee for all $k,l \in \Z^d$.
\end{thm}

\begin{rem}
That the estimate (\ref{eq:dynamicallocalization}) implies  almost sure spectral localization on $(a,b)$ can be shown by means of arguments of Enss-Veselic \cite{EV} on the geometric characterization of bound states. Also, (\ref{eq:dynamicallocalization}) directly prevents the spreading of the wave function over all times, in the sense that for all $p>0$, $\sup_{n\in {\mathbb Z}}\||X|_e^p U_\omega^nP_{[a,b]}^\omega\ffi\|<\infty$ almost surely. Both these facts are explicitly shown in \cite{HJS}.
\end{rem}

Note that specializing to the case $f(z)=z^n$, with $n\in \Z$, we get the localization results  stated in the previous sections. 

\medskip

The proof of Theorem \ref{thm:dynamicallocalization} requires a link between the fractional powers of the resolvent and the resolvent itself, so that some functional calculus can be applied to control operators of the form $f(U)$, for certain continuous functions $f: {\mathbb S}\ra\C$. This is done in two steps. The first one is an estimate on the expectation of the square of the Green function in terms of the expectation of fractional powers of the Green function. This step is equivalent in our unitary framework to the second moment estimate proven by Graf in \cite{Graf} for the self-adjoint case.
\begin{prop} \label{prop:2ndmoment}
Assume B. Then for every $s\in (0,1)$ there exists
$C(s)<\infty$ such that
\be \label{eq:2ndmomentbound}
\E((1-|z|^2)|G(k,l;z)|^2) \le C(s) \sum_{|m-k|\le 4}
\E(|G(m,l;z)|^s) \ee for all $|z|<1$ and $k,l \in \Z^d$.
\end{prop}

\begin{rem}
The fact that the sum in the right hand side of the inequality only carries over indices $m$ a finite distance away from $k$ is a direct consequence of fact that the deterministic operator $S$ has a band structure.
\end{rem}

The second step consists in  
reducing bounds for $f(U)$ to bounds on resolvents by means of the following result.
\begin{lem}
\be \label{f(u)}
f(U)=w-\lim_{r\ra
1^-}\frac{1-r^2}{2\pi}\int_0^{2\pi}(U-re^{i\theta})^{-1}(U^{-1}-re^{-i\theta})^{-1}f(e^{i\theta})d\theta
\ee
for $f\in C({\mathbb S})$ and $U$ a unitary operator. 
\end{lem}
\begin{rem}
This formula is a consequence of the representation of non-negative Borel measures on $\mathbb T$ by Poisson integrals. This can be seen by considering  the non
negative spectral measure $d\mu_\ffi$ on  the torus $\mathbb T$ associated with a normalized $\ffi\in{\mathcal H}$ such that
$
\bra \ffi | U \ffi\ket=\int_{\mathbb T}e^{i\alpha}d\mu_\ffi(\alpha),
$
and 
\be\label{pi}
(1-r^2)\bra \ffi
|(U-re^{i\theta})^{-1}(U^{-1}-re^{-i\theta})^{-1}\ffi\ket=
\int_{\mathbb
T}\frac{1-r^2}{|e^{i\alpha}-re^{i\theta}|^2}d\mu_\ffi(\alpha). 
\ee
For any $f\in C({\mathbb S})$, we thus have
\be
\bra \ffi | f(U)\ffi\ket=\lim_{r\ra
1^-}\int_0^{2\pi}\int_{\mathbb
T}\frac{1-r^2}{|e^{i\alpha}-re^{i\theta}|^2}d\mu_\ffi(\alpha) f(e^{i\theta})\frac{d\theta}{2\pi}
\ee
and one concludes by polarization.
\end{rem}
\medskip

If the fractional moments of the resolvent are exponentially decaying, {\it i.e.} if (\ref{eq:fmexpdecay}) holds, so is the left hand side of (\ref{eq:2ndmomentbound}). Then, considering matrix elements of (\ref{f(u)}) and applying Fatou's lemma and Cauchy Schwarz,  one derives the upper bound (\ref{eq:dynamicallocalization}), as shown in \cite{HJS}.

\bigskip

We have seen that showing dynamical localization for a concrete model amounts to proving that the fractional moments of the resolvent are exponentially decaying, {\it i.e.} that (\ref{eq:fmexpdecay}) holds. This has been done in different ways for the different regimes and models considered. We shall not attempt to explain in details how of this technical task is achieved in the models considered above, but we just want to describe the methods employed to do so.

\medskip

For one dimensional models, either on $l^2(\N)$ or on $l^2(\Z)$, one studies the generalized eigenvectors of the problem, {\it i.e.} the solutions to $U_\omega \psi=z\psi$ in $l(\N)$ or $l(\Z)$. Because of the band structure of the operator $U_\omega$, the generalized eigenvectors are obtained by means of a transfer matrix formalism and their behavior at infinity is controlled by the associated Lyapunov exponent. Exploiting the way the randomness appears in the model, one then shows that the Lyapunov exponent is positive and continuous in the spectral parameter $z$, in a neighborhood of the unit circle. Then, by making use of the expression of the Green function in terms of certain generalized eigenvectors, one shows that (\ref{eq:fmexpdecay}) holds throughout the spectrum, and for all values of the parameter $t\in (0,1)$. This strategy was implemented in \cite{HJS} for the magnetic ring model and for the one dimensional unitary Anderson model, and in \cite{jm} for the quantum walks in random environments models. Previous studies of the properties of the Lyapunov exponents for these models were performed \cite{BHJ, j, hjs}, which lead to spectral localization results by spectral averaging, according to a unitary version of the argument of Simon-Wolff, \cite{SW}.

\medskip

For the $d$-dimensional unitary Anderson model, the large disorder regime was addressed in \cite{j2}. It was shown in this paper that estimate (\ref{eq:fmexpdecay}) holds in any dimension, provided $t$ is small enough. To prove this estimate, the similarity in the way the randomness appears in the model (\ref{uam}) with the discrete Anderson model was used explicitly. The analysis is based on estimates on the expectation of the resolvent equation raised to a fractional power $s$, on rank one perturbation formulas and on a so called "decoupling Lemma", similar to the one shown in \cite{AM} for the self-adjoint case. This leads to an inequality satisfied by the function $0\leq f(k)=\E(|F(k,j;z)|^s)$ in $l^\infty(\Z^d)$, where $F(z)=U_\omega(U_\omega-z)^{-1}=\I+z(U_\omega-z)^{-1}$ is essentially equivalent to the resolvent. This inequality says that $f(k)$ is smaller than a $z$-independent constant times the weighted average of its values around $k$, with weights given by the matrix elements of $S_d$. The structure of $S_d$ and dependence in $t$ of its matrix elements then imply the sought for bound, for $t$ small enough.
 
 \medskip
 
 The band edge regime for the $d$-dimensional unitary Anderson model was tackled in the paper \cite{HJS}, adapting the general strategy provided in \cite{AENSS}. This  regime, which is the most challenging to cover, requires getting finite volume estimates on the resolvent, close to the band edges. A first step consists in defining the restriction $U_\omega |_{\Lambda(L)}$ of $U_\omega$ to finite boxes ${\Lambda(L)}\subset \Z^d$ of side length $L$ by means of appropriate boundary conditions which make this restriction unitary and imply certain monotony properties of the spectrum as boxes are spit by adding more boundary conditions. Then, one needs to get accurate probabilistic bounds on the size of the resolvent of this restriction, when the spectral parameter $z$ is close to the band edges. It requires showing that when $L$ becomes large, the probability to have eigenvalues a distance smaller than $1/L^\beta$ away from the band edges is of order $e^{-\gamma L^\alpha}$, for $0<\beta<1$ and $\alpha, \gamma >0$, {\it i.e.} a Lifshitz tail type estimate. Then a decoupling lemma with an iterative argument allows us to prove the bound (\ref{eq:fmexpdecay}) for the infinite volume operator $U_\omega$, in  a non-empty neighborhood of the band edges.

\medskip

Finally, we would like to mention that there is at least one more popular model in condensed matter physics whose dynamics reduces to the study of a discrete time quantum dynamics generated by a random unitary operator with a band structure: the Chalker Coddington model and its variants, see \cite {cc}. This model can be thought of as a unitary equivalent of the discrete Schr\"odinger equation on a finite width two-dimensional strip. Some progress was made recently about the properties of this model in \cite{abj}. But the focus of this work is more on the analysis of the associated set of Lyapunov exponents than on dynamical localization aspects. This is why we didn't provide a description of the Chalker Coddington model in these notes, eventhough it certainly belongs to the family of unitary random operators presented here.

\medskip

{\bf Acknowledgements: } It is a pleasure to thank Bob Sims and Daniel Ueltschi for the invitation to the perfectly organized "Arizona School of Analysis with Applications 2010", where part of this material was presented.

\end{document}